\documentstyle[twocolumn,floats,aps]{revtex}

\begin{document}
\draft

\twocolumn[\hsize\textwidth%
\columnwidth\hsize\csname@twocolumnfalse\endcsname

\title{\bf Where is the Luttinger liquid in one dimensional semiconductor
quantum wire structures?}

\author{D.W. Wang$^{(1)}$, A. J. Millis$^{(2)}$, and S. Das Sarma$^{(1)}$}

\address{
(1)Department of Physics, University of Maryland,
College Park, Maryland 20742-4111\\
(2)Center for Materials Theory and Department of Physics and 
Astronomy, Rutgers University, New Brunswick, New Jersey 08554\\
}

\date{\today}
\maketitle
\pagenumbering{arabic}


\begin{abstract}
We present the theoretical basis for analyzing resonant Raman scattering
experiments in one-dimensional systems described by the Luttinger liquid
fixed point. We make experimentally testable predictions for distinguishing
Luttinger liquids from Fermi liquid and argue that presently available
quantum wire systems are \textit{not} in the regime 
where Luttinger liquid effects are important.
\end{abstract}

\vskip 1pc]
\narrowtext
It is theoretically well-established [1-3] that a one dimensional
interacting electron system (1DES), is \textit{not} a Fermi liquid
(FL). Unlike a Fermi liquid, the interacting 1DES has neither sharp fermionic
quasiparticle excitations nor a discontinuity in the electron momentum
distribution function. The elementary excitations are charge $e$, spin zero
bosons and spin 1/2 charge 0 'semions' (fractional statistics objects), and
the fermion is a composite of these. Interacting 1DES have been generically
termed Luttinger Liquids (LL) [2] and have been the subject of
extensive theoretical study over the last 40 years and particularly over the
last decade. Despite the intense theoretical interest, there have been few
convincing experimental demonstrations of the predicted LL behavior in real
1DES. The power-law density of states observed in tunneling into edges of
quantized hall systems [4] have been interpreted in terms of the
theoretically expected 'chiral Luttinger liquid' behavior of edge states.
The origin of the differences between the observed and expected exponents
is presently an area of active inquiry. Photoemission experiments on Mott
insulating oxides have been interpreted in terms of the 'holon' and 'spinon'
excitations of a charged Luttinger liquid [5]. 

A 1DES which is of particular interest both for fundamental physics and for
technology is the system formed in GaAs-based semiconductor quantum wire
(QWR) structures. Modern materials growth and fabrication techniques have
produced nearly ideal 1DES in which the electron may move freely only along
the length of the wire. The transverse motion is quantized with the quantum
1D subbands separated by several meV. It is possible to 
have low enough carrier densities
so that at low temperatures only the lowest 1D subband is occupied
by electrons. Such GaAs QWR based 1DES should be ideal systems for the study of
interacting electrons in one dimension because they are free from
complications arising from band structure, lattice effects, and crossovers
to three dimensional behavior which often make interpretations of
experimental data difficult in more traditional 1DES based on organic
compounds.

It is surprising, therefore, that no definitive LL behavior has been
reported in GaAs QWR systems, and in fact the 1D Fermi gas/liquid model
seems to ''work'' operationally very well in describing and explaining the
observed 1DES experimental properties in GaAs QWR [6,7]. 
Part of the reason for the apparent absence of the expected LL
behavior is undoubtedly the fact that in weakly interacting 1DES, \textit{at
finite temperatures and in the presence of impurity scattering}, the actual
quantitative difference between a LL and a FL is not large [7], although the
qualitative conceptual difference between the two is huge. A more
fundamental issue is that the differences between a Luttinger Liquid and a
Fermi liquid are most obvious in the one-electron spectrum, while the
experimental probes which may most conveniently be applied to the QWR
structures produce 'particle-hole pairs'. The differences in particle-hole
pair properties between Luttinger Liquid and Fermi Liquid systems is much
less pronounced than are the differences in the one electron spectrum. This
perhaps accounts for the fact that one of the most important probes of QWR
structures, resonant inelastic light scattering or Raman scattering
spectroscopy (RRS) [6,8], has not yet observed any definitive indications of
LL behavior in these systems.

In RRS experiments, light is absorbed at one frequency and re-emitted at
another, creating one or more particle-hole pairs. In the so-called
polarized geometry with the incident and outgoing photons having the same
polarization (so that no spin is transferred to the QWR), RRS experiments in
GaAs QWRs consistently [6,8] show two peaks which indeed look qualitatively
very similar [9] to the spectra for the corresponding 2D and 3D systems. In
these higher dimensional systems, the two peaks have a clear and generally
accepted Fermi liquid interpretation [9]. The higher energy peak is associated
with the plasmon or charge density excitation (CDE), a collective density
excitation of the electron gas, and the lower energy spectral peak is
associated with incoherent particle-hole pair excitations (SPE). In the
QWR materials, the lower energy peak occurs at an (approximate) excitation
energy of $\sim qv_{F}$, where $q$ is the excitation momentum and $v_{F}$ is
the 1D Fermi velocity obtained from the band structure of the QWR. An
interpretation of the lower peak as an SPE contribution seems therefore
natural [9]. However, there is a strong theoretical objection to this
interpretation: in a one dimensional system there is spin-charge
separation: the only charge excitations live at the plasmon frequency, and
cannot contribute to excitations at the SPE energy. The signal observed in
this $q,\nu$ range must be due to the chargeless spin excitations
of the LL; in
particular it is possible to combine two $S=1/2$ excitations into a $S=0$
object, creation of which is allowed by the Raman selection rules. 

Sassetti and Kramer (S-K) presented a qualitative theory of this effect [10]. 
They showed that although the leading contribution to the RRS matrix element
corresponds to coupling the light to the electron density operator, there is
a sub-leading term (which becomes more important under resonance conditions)
which may be interpreted as a coupling of light to the energy density
fluctuations of the electrons in the QWR.  The energy density fluctuations
have a contribution from the spin excitations, which qualitatively explains
the data, but the S-K theory did not calculate the spectral weights of the 
RRS peaks. Too close to resonance, the S-K theory breaks down. The S-K work
also does not show how to distinguish a LL from a FL in the RRS experiment.
The most important theoretical problem is that the S-K calculation is
logically inconsistent, because it uses an expression for the RRS matrix
element which is correct only if the conduction band is a FL not an LL. Thus
S-K uses FL matrix elements but LL excitations. In
our paper the correct LL matrix element is used, leading to expressions
different from those derived by S-K.

In this paper we present an essentially complete treatment of RRS in a one
dimensional electron gas. We obtain a precise expression for the energy
transferred to the QWR in a RRS experiment, valid at all values of the
difference of the energy from resonance, and evaluate it quantitatively in
several experimentally relevant limits. We show which features of the data
contain information about the LL exponents, obtain expressions for the
relative amplitudes of the SPE and CDE peaks, determine lineshapes and
discuss qualitatively the crossover from LL to FL behavior. 

Resonant Raman scattering is a two-photon process in which a photon is
absorbed, transferring an electron from the valence $(V)$ band to the
conduction $(c)$ band and a photon is emitted, transferring an electron from
the conduction band back to the valence band. We assume that the valence
band is initially filled, and assume there is no excitonic interaction
between conduction and valence band states. The excited valence hole is then
described by a single-particle Hamiltonian, which we write as $H_{V}$, while
the conduction band is described by some interacting Hamiltonian which we
denote $H_{LL}$. We denote the photon absorption and emission by $P_{1,2}$
respectively.
The RRS process is described by the following Hamiltonian:
\begin{equation}
H=H_{V}+H_{LL}+\widehat{P}_{1}+\widehat{P}_{2}
\end{equation}
where the photon-in $(P_{1})$ and photon-out $(P_{2})$ terms are
\begin{eqnarray}
\widehat{P}_{1}&=&e^{-i(\Omega +\nu /2 )t}\sum_{p,s}
c_{p+q/2,s}^{\dagger}(t)v_{p,s}(t)+\mathrm{h.c.} \\
\widehat{P}_{2}&=&e^{i(\Omega -\nu /2)t}\sum_{p,s}
v_{p,s}^{\dagger}(t)c_{p-q/2,s}(t)+\mathrm{h.c.}
\end{eqnarray}
with $c$ and $v$ the annihilation operators for electrons in conduction and
valence band states respectively. Note that
the operator $v_{p,\sigma }^{\dagger}$ creates an eigenstate of $H_{V}$
with energy $E_{p}^{V}$ while the $c_{p,\sigma}^{\dagger}$ 
operators does \textit{not} create eigenstates of
$H_{LL}$. The absorbed(emitted) photon energy and momentum are set
$\Omega\pm\nu/2$ and $\pm q/2$ respectively.

We now use the standard methods of time-dependent perturbation theory to 
calculate the amplitude, $a_n(t_0)$, for the system at some time $t_0$
to be in some excited state $|n\rangle$ of QWR, but with no holes in the 
valence band. We assume the system is in its ground state at $t=0$. 
Our neglect of any excitonic interaction between conduction and valence 
band simplifies the calculation and we obtain
\begin{equation}
a_n(t_0)=\frac{1}{L}\sum_{r,s}\int dRe^{-iqR}\int_{0}^{t_0} dTe^{i\nu T}
\langle n|\widehat{O}_{rs}(R,T)|0\rangle
\end{equation}
with
\begin{eqnarray}
&&\widehat{O}_{rs}(R,T)=\int dx\int_0^{T}dt\,\phi
(x,t)\times\nonumber\\
&&\hspace{1.cm}\psi_{rs}(R+x/2,T+t/2)\psi_{rs}^{\dagger}(R-x/2,T-t/2),
\end{eqnarray}
where $r$ and $s$ are band and spin indices ($pm 1$), and
\begin{equation}
\phi (x,t)=e^{i\Omega t}\sum_{p}e^{i(E_{p}^{V}t-px)}.
\end{equation}

Eqs. (4) and (5) are our fundamental new results: 
they show that the RRS process 
acts to create a particle-hole pair at a spatial separation $x$ and 
temporal separation $t$. These are determined by the average photon 
frequency $\Omega$ and the valence-band properties encoded in $E_p^V$. 
Further, if interactions are present in the conduction band, the states 
created by $\psi^\dagger$ and by $\psi$ are not eigenstates of $H_{LL}$ 
and therefore the matrix element is itself modified by interactions. 

We note that Eqs. (4) and (5) maybe 
substantially simplified in the limit of greatest physical interest. We 
linearize the valence band energy about the conduction band Fermi momentum, 
writing $E^V_F=-\Delta-v_F^V(rp-p_F)$ for branch $r$ and define 
$\omega_R=\Omega-\Delta$ as the photon frequency with respect to the resonance
energy, $\Delta$. The $p-$integral gives $\delta(x+v_F^Vt)$. 
Finally we write the conduction band operators in terms of the bosons 
which create eigenstates of $H_{LL}$, and normal-order in the boson 
basis, obtaining
\begin{figure}

 \vbox to 10cm {\vss\hbox to 5.cm
 {\hss\
   {\includegraphics{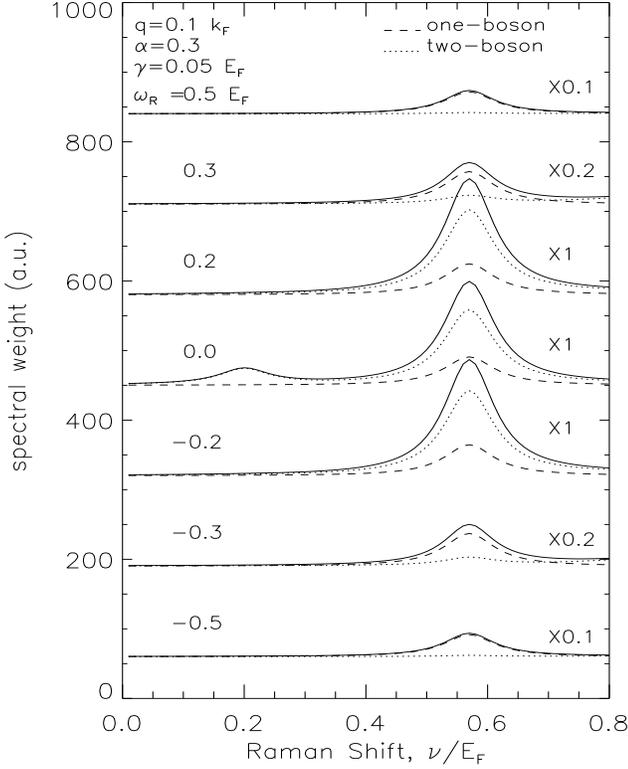}
   }
  \hss}
 }
\caption{
Calculated polarized RRS spectra for various resonance condition, $\omega_R$. 
One- and two-boson contributions have been plotted separately in order to show
their relative contributions (see text). A finite broadening $\gamma$
has been used to depict the results.
Note that the overall spectral weights decreases dramatically off-resonance,
as indicated by the individual scale factors on right side of each plot.
}
\end{figure}
\begin{figure}

 \vbox to 5.5cm {\vss\hbox to 5.cm
 {\hss\
   {\includegraphics{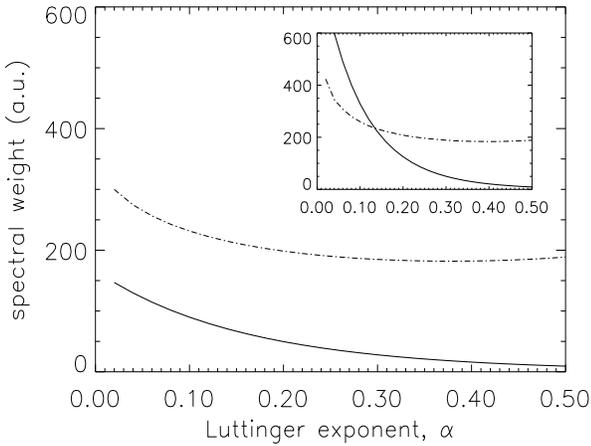}
   }
  \hss}
 }
\caption{
Spectral weights for the low energy (solid curve) and the high energy (dashed)
RRS peaks in the LL theory plotted A  function of the Luttinger exponent 
$\alpha$: $\omega_R=0$ (main); 0.1 (inset). When $|\omega_R|$ larger than 0.1
the low energy ("SPE") weights are always much smaller than the high energy
(CDE) weights over the whole range of $\alpha$. 
}
\end{figure}
\begin{eqnarray}
\widehat{O}_{rs}(R,T)&=&L\int_{0}^{T}dt
\,e^{i\omega_Rt}G^c_{rs}(-rv_F^Vt,t) \nonumber\\
&&\hspace{-0.5cm}:e^{i\Phi _{rs,\rho}(R,-rv_{F}^Vt;T,t)}:
:e^{i\Phi _{rs,\sigma}(R,-rv_{F}^Vt;T,t)}:,
\end{eqnarray}
where
\begin{eqnarray}
&&\Phi _{rs,\rho}(R,x;T,t)=2\sum_{p>0}e^{-\alpha p/2}\sqrt{\frac{
\pi }{pL}}\times  \nonumber \\
&&\left\{-\sinh\theta_\rho\sin [p(rx+v_\rho t)/2]
[b_{-rp}^{\dagger}e^{ip(rR+v_\rho T)}+\mathrm{h.c.}]
\right.\nonumber \\
&&+\left.\cosh\theta_\rho\sin [p(rx-v_\rho t)/2][b_{rp}
^{\dagger}e^{-ip(rR-v_\rho T)}+\mathrm{h.c.}]\right\},
\nonumber\\&& \\
&&\Phi _{rs,\sigma}(R,x;T,t)=2s\sum_{p>0}e^{-\alpha p/2}
\sqrt{\frac{\pi }{pL}}\times  \nonumber \\
&&\left\{ \sin [p(rx-v_{F}^ct)/2]
[\sigma_{rp}^{\dagger}e^{-ip(rR-v_{F}^{c}T)}+\mathrm{h.c.}]\right\}.
\end{eqnarray}
Here $b^{+}$ and $\sigma^{+}$ create charge and spin excitations respectively
and $v_\rho=v_F^c e^{-2\theta_\rho}$ is the plasmon velocity, where
the exponent $e^{-2\theta_\rho}=\sqrt{1+2g/\pi v^c_F}$ is defined
for the short-ranged interaction, $g$.
$G^c$ is the exact conduction band Green's function at spatial 
separation $-rv_F^Vt$, and temporal separation $t$.
We have assumed that the interactions are negligible in the spin sector
and therefore the spin excitation velocity is just the Fermi velocity.
As long as $v_F^V$, 
the valence band velocity at the conduction band $p_F$ is different from 
the spin and charge velocities of Luttinger liquid, $G^c$ is a 
decaying function of $t$. In the noninteracting case, $G^c\sim 1/t$; 
interaction leads to a faster decay: $G^c\sim 1/t^{1+\alpha}$ with the LL
exponent $\alpha=\sinh^2\theta_\rho >0$ (not the same one as we use in Eqs.
(8-9) for infinitely small convergent factor) for short-ranged 
interactions; $G^c$ decays faster with the
physically relevant long-ranged interactions. This faster decay of $G^c$ 
is the mathematical expression of the renormalization of the RRS vertex 
by the interactions, which produce the Luttinger liquid behavior. 
As we will now show, it has 
important consequences for various aspects of the RRS spectra; and in 
particular for the dependence of the CDE and SPE energies on the 
difference of the average photon energy from the resonance.

We defer to a subsequent paper a full evaluation of the RRS correlation 
function, which is computationally demanding and not very illuminating, 
and present here the results of expanding Eq. (7) in terms of 
boson operators. The essential point is that if the combination of 
$e^{i\omega_Rt}G^c(-rv_F^vt,t)$ decays rapidly as $t$ increases (large
$\omega_R$ as off-resonance or large $\alpha$ as strong interaction), then 
the $t-$integral is dominated by small times and an expression in 
power of bosons is rapidly convergent. We will show below that the first order
term, one-boson result, gives the main contribution to CDE spectrum and
dominates the whole spectrum as off-resonance and the second order term, 
two-boson (spinon) result, gives the peak at "SPE" energy as near resonance, but
it still has relatively small weights as compared to the first order CDE.

Expanding the exponentials, keeping only the one-boson term and integrating
explicitly, gives the one-boson transition rate as a delta function at 
$\nu=qv_\rho$ with the spectral weight ($\alpha<1$)
\begin{equation}
W_{1}=\frac{2L\Gamma^2(-\alpha)}{qv_\rho^2}
\left|\left(\frac{\omega_R-\omega_q}{E_0}\right)^\alpha-
\left(\frac{\omega_R+\omega_q}{E_0}\right)^\alpha
\right|^2,
\end{equation}
where $\omega_q\equiv qv_\rho/2$, neglecting $v_F^V$ for simplicity. $E_0$
is the energy scale depending on the interaction range and roughly
of the order of Fermi energy, $E^c_F$. 
For $\omega_q\ll |\omega_R|$, $W_1\propto |\omega_R|^{2\alpha-2}$,
while for $\omega_R=0$, $W_1\propto \sin^{2}(\pi \alpha /2)$.
Thus LL effects enter the CDE portion (one-boson) of the spectrum
in two ways (for short-ranged interaction): first,
far from resonance, it changes the frequency dependence of spectral weights
from $\omega_R^{-2}$, the noninteracting result, to $\omega_R^{-2+2\alpha }$
(note that all other higher order bosonic contribution decays much faster, this
confirms the validity of the bosonic expansion we mentioned above).
and secondly as on resonance ($\omega_R=0$) it changes the value to 
be nonzero due to finite interaction strength.

To second order, two new effects appear. In the density spectrum, branch mixing
process appear. These lead to a continuum absorption beginning at the CDE 
threshold, $\omega=qv_\rho$. In addition, an $S=0$ combination of 
spin excitations may be excited via the two spinon, 
$\langle\sigma\sigma,\sigma\sigma\rangle$ (note that there is \textit{no}
first order contribution in spin channel due to the selection rule
of RRS in the polarized spectroscopy), and gives the so-called "SPE" mode
at $\nu=qv^c_F$. 

In Fig. 1, we show the spectrum from one and two bosons in different resonance
energy. One can find that (i) the overall spectral weights decays very fast
off resonance, and (ii) the "SPE" peak is generated at $\omega\sim 0.2E^c_F$
by the two-boson contribution near resonance. But as compared with the
CDE peak at plasmon energy (about 0.57 $E^c_F$), the "SPE" peak is still very
small compared with CDE. This striking result arises from the fact that
the contribution of one spin-boson in the first order is forbidden by the
specific selection rule of polarization in depolarized RRS spectroscopy. (iii)
At higher energy side above CDE peak, there is some continuum structure which
is not shown in the range of Fig. 1. This continuum is from the interaction
between different branches of charge bosons due to finite $g_2$ interaction.
We are not interested in their structure because it goes to zero near the
plasmon energy and their higher energy behavior is off the
experimentally measurable region, and become unphysical due to the failure of
the linear dispersion assumption. (iv) When
including three or higher order boson contribution (not shown in this paper), 
we will see the mixture of charge boson and spin boson in a form like
$\langle\sigma\sigma\rho,\rho\sigma\sigma\rangle$, which will come into the
energy between $qv_F^c$ and $qv_\rho$, plasmon energy, as a continuum 
structure.
A detailed analysis shows that these higher order contribution is 
relatively small and no special structure compared to the first two order 
result we present here. 
While Fig. 1 is for a specific value of $\alpha$ ($=0.3$) we show in Fig. 2
the calculated charge boson and spin boson RRS spectral weights at 
resonance and away from resonance. In general, the LL theory predicts much
smaller spectral weight for the lower energy "SPE" mode than the FL theory 
[9] at resonance. This is particularly true since our best estimate 
for the Luttinger exponent of the experimental system [6] 
(obtained from the CDE energy dispersion) is $\alpha\sim 0.4$.

As compared with the experimental result, which shows possible comparable 
spectral weight of "SPE" with CDE [6], we find that the LL theory 
result induced
by resonance effect does \textit{not} explain the experimental results
quantitatively, even though we could recover the SPE peak through the coupling
of two spinon in LL, not as the SPE in FL theory. This inconsistency cannot be 
resolved even evaluating the full bosonic contribution without expansion
as what we have in this paper. Our future work shows that the spectral weight 
of the "SPE" peak enhanced by spinon coupling in polarized RRS spectra is 
always relatively small compared to that of CDE. Therefore, unlike the 
conclusion of previous work [10] based on the incorrect matrix element, 
we claim that the
whole problem cannot be simply understood by the correct LL theory.
We believe that the existing experimental results [6] are in the high 
energy crossover regime where in fact a FL description maybe more 
appropriate for the RRS data than the LL description which is an 
asymptotic low energy description. This explains the spectacular 
quantitative success of the FL RRS theory developed in ref. [9].

In conclusion, we provide the correct LL theory for the RRS
spectra calculation, and obtain some meaningful and interesting results 
to study the possible origin of LL features in the RRS spectra of 1D
QWR systems. We also develop an useful bosonic expansion method to
study the two-particle correlation function. Finally, we find that the LL
theory cannot quantitatively explain the experimental data most likely because
the RRS experiments are not in the asymptotic low energy LL regime.

\end{document}